\documentclass[10pt,journal]{IEEEtranTCOM}

\usepackage[english]{babel}
\usepackage[usenames]{color}
\usepackage[cp1250]{inputenc}
\usepackage{amsfonts}
\usepackage{amsthm}
\usepackage{graphicx}
\usepackage{epsfig}
\usepackage{mathrsfs}
\usepackage{amsmath}
\usepackage{algorithm}
\usepackage{algorithmic}
\usepackage{hyperref}

\theoremstyle{plain}

\begin{document}
\newcommand{\bea}{\begin{eqnarray}}
\newcommand{\eea}{\end{eqnarray}}
\newcommand{\be}{\begin{equation}}
\newcommand{\ee}{\end{equation}}
\newcommand{\beas}{\begin{eqnarray*}}
\newcommand{\eeas}{\end{eqnarray*}}
\newcommand{\bs}{\backslash}
\newcommand{\bc}{\begin{center}}
\newcommand{\ec}{\end{center}}
\def\SC {\mathscr{C}}

\title{Improving distribution and flexible quantization\\ for DCT coefficients}
\author{\IEEEauthorblockN{Jarek Duda}\\
\IEEEauthorblockA{Jagiellonian University,
Golebia 24, 31-007 Krakow, Poland,
Email: \emph{dudajar@gmail.com}}}
\maketitle
\begin{abstract}
While it is a common knowledge that AC coefficients of Fourier-related transforms, like DCT-II of JPEG image compression, are from Laplace distribution, there was tested more general EPD (exponential power distribution) $\rho\sim \exp(-(|x-\mu|/\sigma)^{\kappa})$ family, leading to maximum likelihood estimated (MLE) $\kappa\approx 0.5$ instead of Laplace distribution $\kappa=1$ - such replacement gives $\approx 0.1$ bits/value mean savings (per pixel for grayscale, up to $3\times$ for RGB).

There is also discussed predicting distributions (as $\mu, \sigma, \kappa$ parameters) for DCT coefficients from already decoded coefficients in the current and neighboring DCT blocks. Predicting values $(\mu)$ from neighboring blocks allows to reduce blocking artifacts, also improve compression ratio - for which prediction of uncertainty/width $\sigma$ alone provides much larger $\approx 0.5$ bits/value mean savings opportunity (often neglected).

Especially for such continuous distributions, there is also discussed quantization approach through optimized continuous \emph{quantization density function} $q$, which inverse CDF (cumulative distribution function) $Q$ on regular lattice $\{Q^{-1}((i-1/2)/N):i=1\ldots N\}$ gives quantization nodes - allowing for flexible inexpensive choice of optimized (non-uniform) quantization - of varying size $N$, with rate-distortion control. Optimizing $q$ for distortion alone leads to significant improvement, however, at cost of increased entropy due to more uniform distribution. Optimizing both turns out leading to nearly uniform quantization here, with automatized tail handling.

\end{abstract}
\section{Introduction}
Modern lossy image/video compression is usually based on Fourier-related transforms, especially discrete cosine transform DCT-II used e.g. in JPEG image compression~\cite{jpeg}. While DC coefficients describing mean value have completely different behavior, requiring separate treatment usually similar to lossless image compression, the AC coefficients are usually assumed to be from Laplace distribution~\cite{jq}.

\begin{figure}[t!]
    \centering
        \includegraphics{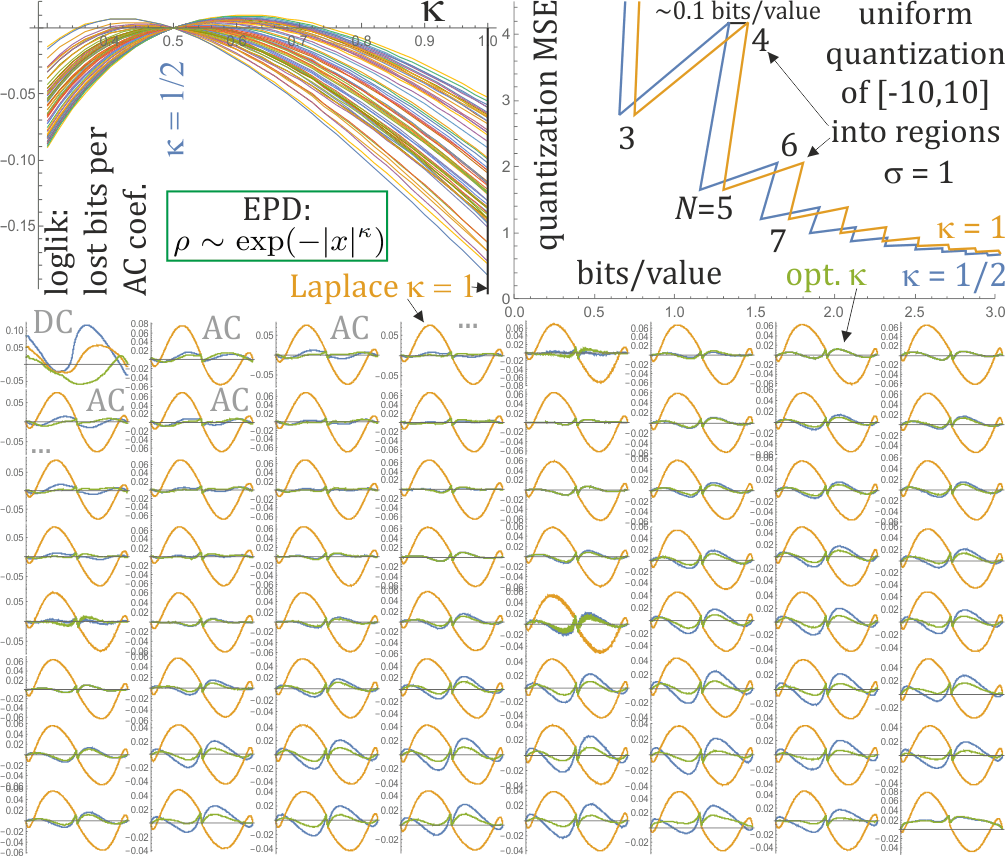}
        \caption{Evaluation using 48 grayscale 8bit 512x512 images (values normalized to $[0,1]$) from http://decsai.ugr.es/cvg/CG/base.htm . \textbf{Top left}: log-likelihood evaluation (mean $\lg(\rho(x))$) for 63 AC coefficients of 8x8 DCT-II transform for EPD family $\rho(x)\propto \exp(-(|x-\mu|/\sigma)^\kappa/\kappa)$, shifted to zero for $\kappa=1/2$ (remaining $\sigma, \mu$ parameters from MLE). Vertical difference can be interpreted as change in bits/pixel for using different $\kappa$: we can see that going from $\kappa=1$ of Laplace distribution to $\kappa=1/2$, we get $\approx 0.11$ bits/value mean savings, further individual $\kappa$ optimization gave additional $\approx 0.03$ bits/value mean savings. \textbf{Top right}: rate-distortion comparison for size $N$ uniform quantization (tails go to extremal nodes) on $[-10,10]$ range of $\kappa=1/2, \sigma=1$, $\mu=0$ EPD distribution using this density (blue) or $\kappa=1$ standard Laplace assumption (orange) - we can see these $\approx 0.1$ bits/value savings from switching to $\kappa=1/2$, nearly universal for various quantization size $N$. Valuable observation is that quantization into even $N$ is significantly worse - should be avoided, focusing on odd $N$. \textbf{Bottom}: evaluation of various distributions for 8x8 DCT coefficients - perfect agreement would have flat line in zero. Specifically, the values were transformed $y=CDF(x)$ using CDF of assumed distribution, then sorted (empirical distribution function) should ideally give diagonal - which is subtracted. We can see that Laplace has relatively large $0.04-0.08$ disagreement (orange), it is much smaller for $\kappa=1/2$ (blue), sometimes a bit further improved for individually optimized $\kappa$. Visually their main imperfection is large jump in the center: corresponding to increased probability of zero value, what can be included in probabilities used for quantization.}
       \label{epd}
\end{figure}

This assumption is verified here using more general family: EPD (exponential power distribution)~\cite{exp,exp1}: $\rho(x)\propto \exp(-(|x-\mu|/\sigma)^\kappa/\kappa)$ covering both Laplace distribution for $\kappa=1$, Gaussian distribution for $\kappa=2$, and other behaviors of both body and tail of distribution. MLE (maximum likelihood estimation) allows to test if standard $\kappa=1$ assumption is the proper one, but for AC coefficients it clearly leads to essentially smaller $\kappa\approx 1/2$, as shown if Fig. \ref{epd}.

Such replacement allows to improve compression ratio by $\approx 0.1$ bits/value, which seems significant as for RGB we get $\approx 0.3$ bits/pixel this way (or less for chroma subsampling), could allow for better rate-distortion control, or other optimizations e.g. of PVQ (perceptual vector quantization)~\cite{pvq,pvq1}.

While DCT transform decorrelates data in a block, there remain other statistical dependencies like homoscedasticity - discussed predicting width (scale parameter) for AC coefficients based on already decoded coefficients: as linear combination of their absolute values, leads to $\approx 0.5$ bits/value mean savings. There is also discussed predicting coefficients and their uncertainty from already decoded neighboring blocks - reducing blocking artifacts and further improving compression ratio.

There is also discussed inexpensive automatic approach for quantization especially of such continuous probability distribution functions - by first optimizing quantization density function $q$ describing how choose local density of quantization nodes for asymptotic case of infinite number of region, then use it for finite number of regions. Specifically, $q$ integrates to 1 as density, its inverse CDF (cumulative distribution function) on regular lattice of chosen size $N$ gives the quantization points.

For minimizing distortion for given density $\rho$ - usually MSE (mean squared error) of quantization, like classical Lloyd-Max algorithm~\cite{lloyd,max}, here we get $q\propto \rho^{1/3}$: that denser regions should have denser quantization, but only with cube root, e.g. twice denser for 8 times larger density.

However, while such quantization indeed reduces distortion, turns out it also increases entropy by more uniform distribution among quantization regions. Optimizing both rate and distortion, such optimization has lead to nearly uniform quantization (at least for such first considered examples) - with optimized tail handling.

This is work in progress, continuation of author's revisitions of basic approaches for image/video compression~\cite{param,param1}, for example for context dependent probability distribution models - what is planned to be explored for DCT coefficients in later versions of this article, alongside other expansions.

\section{Exponential power distribution (EPD)}
\begin{figure}[t!]
    \centering
        \includegraphics{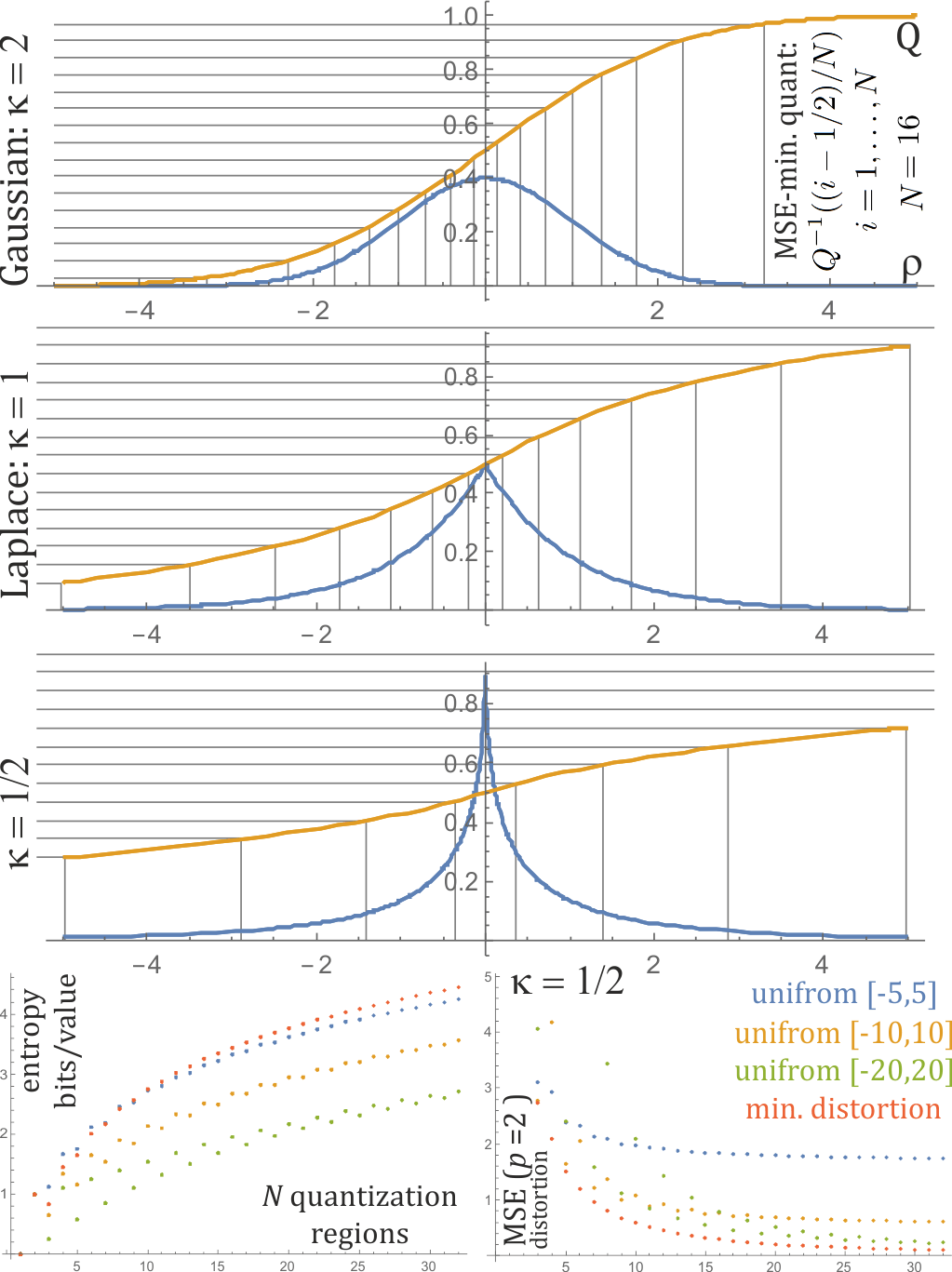}
        \caption{\textbf{Top 3}: Quantization densities minimizing distortion as mean squared quantization error for $\kappa=2$ (Gaussian), $1$ (Laplace) and $1/2$ EPD distributions. As discussed, quantization density is $q\propto \rho^{1/3}$ normalized to $\int q dx=1$, for which we find CDF: $Q(y)=\int_{-\infty}^{y} q(x) dx$ and quantization is given by $Q^{-1}$ on size $N$ regular lattice: $\{(i-1/2)/N:i=1,\ldots,N\}$. \textbf{Bottom}: comparison of entropy and MSE for $N=1,\ldots,32$ and such minimal distortion quantization (red) and 3 uniform quantizations on $[-5,5],[-10,10],[-20,20]$ ranges (the two extremal quantization regions include tails). While we can see that MSE is essentially better, unfortunately it comes with increased entropy (more uniform distribution over quantization regions), not providing clear improvement for common rate-distortion optimization.}
       \label{mindist}
\end{figure}

For $\kappa>0$ shape parameter, $\sigma>0$ scale parameter and $\mu\in\mathbb{R}$ location, probability distribution function (PDF, $\rho_{\kappa\mu \sigma}$) and cumulative distribution function (CDF, $F_{\kappa\mu \sigma}(x)=\int_{-\infty}^x \rho_{\kappa\mu \sigma}(y)dy$) of EPD are correspondingly:
\be\rho_{\kappa\mu \sigma}(x)= \frac{C_\kappa}{\sigma} e^{-\frac{1}{\kappa}\left(\frac{|x-\mu|}{\sigma}\right)^\kappa}\quad\textrm{for}\quad C_\kappa=\frac{\kappa^{-1/\kappa}}{ 2\Gamma(1+1/\kappa)}\label{EPDrho}\ee
$$F_{\kappa\mu \sigma}(x) = \begin{cases}
      \frac{1}{2}\gamma\left(\frac{1}{\kappa},\frac{(|x-\mu|/\sigma)^\kappa}{\kappa}\right) & \text{if}\ x<\mu \\
      1-\frac{1}{2}\gamma\left(\frac{1}{\kappa},\frac{(|x-\mu|/\sigma)^\kappa}{\kappa} \right) & \text{if}\ x\geq \mu
    \end{cases}$$

where $\Gamma$ is Euler gamma function, $\gamma(a,z)=\Gamma(a,z)/\Gamma(a)$ is regularized incomplete gamma function. Their PDFs for $\kappa=1/2,1,2$ are plotted in \ref{mindist}.

Its variance is
\be \textrm{var}=\int_{-\infty}^{\infty} x^2 \rho_{\kappa\mu \sigma}(x) dx = \frac{\kappa^{2/\kappa}\Gamma(3/\kappa)}{\Gamma(1/\kappa)}\ \sigma^2 \ee
which is $\sigma^2$ multiplied by constant decreasing with $\kappa$, e.g. 7.5 for $\kappa=1/2$, 2 for $\kappa=1$ (Laplace distribution), 1 for $\kappa=2$ (Gaussian distribution).

Its (base 2) differential entropy is
\be H=-\int \rho \lg(\rho) = \frac{1}{\kappa \ln(2)} - \lg\left(\frac{\kappa^{1-1/\kappa}}{2\Gamma(1/\kappa)}\right)+\lg(\sigma)  \ee
For uniform quantization $\hat{x}=\textrm{round}(x/q)$, $\tilde{x}=q\,\hat{x}$ with $q$ step lattice: $q\mathbb{Z}$, to store such values we need $h\approx H-\lg(q)$ bits/value. For large $q$ it smoothens to lower bound in 0, allowing to use e.g. $h\approx \ln(\exp(2(H-\lg(q)))+1)/2$ type approximation. For $\kappa=1$ Laplace distribution we get geometric series, allowing for analytical formula $(\Delta=q/\sigma)$:

\begin{small}
$$h=2\frac{\left({\Delta}\,e^{\Delta}  \lg(e) -(e^{\Delta}-1)\lg\left(\sinh\left(\frac{{\Delta}}{2}\right)\right)\right)\sinh\left(\frac{{\Delta}}{2}\right)}{(e^{\Delta}-1)^2}-$$
$$-\left(1-e^{-\frac{{\Delta}}{2}}\right)\lg\left(1-e^{-\frac{{\Delta}}{2}}\right)$$
\end{small}
Distortion as MSE of such uniform quantization
$$MSE(\rho,q)=\sum_{\hat{x}\in \mathbb{Z}} \int_{(\hat{x}-1/2)q}^{(\hat{x}+1/2)q} (x-\tilde{x}q)^2\, \rho(x)\,dx $$
has known analytical formula for $\kappa=1$ (Laplace~\cite{jq}):
\be MSE=\sigma\left(2\sigma-\frac{q}{\sinh(q/(2\sigma))}\right)=
\frac{q^2}{12}-\frac{7q^4}{2880\sigma^2}+\ldots \label{lapq}\ee
Generally, for $q\to 0$ it is $q^2/12$ as we get infinitesimal ranges for which $\int_{-q/2}^{q/2} x^2\, dx =q^2/12$. In contrast, for $q\to \infty$ we approximate all values with (minimizing MSE) expected value 0 here, hence the quantization error approaches variance of the distribution.

\begin{figure}[t!]
    \centering
        \includegraphics{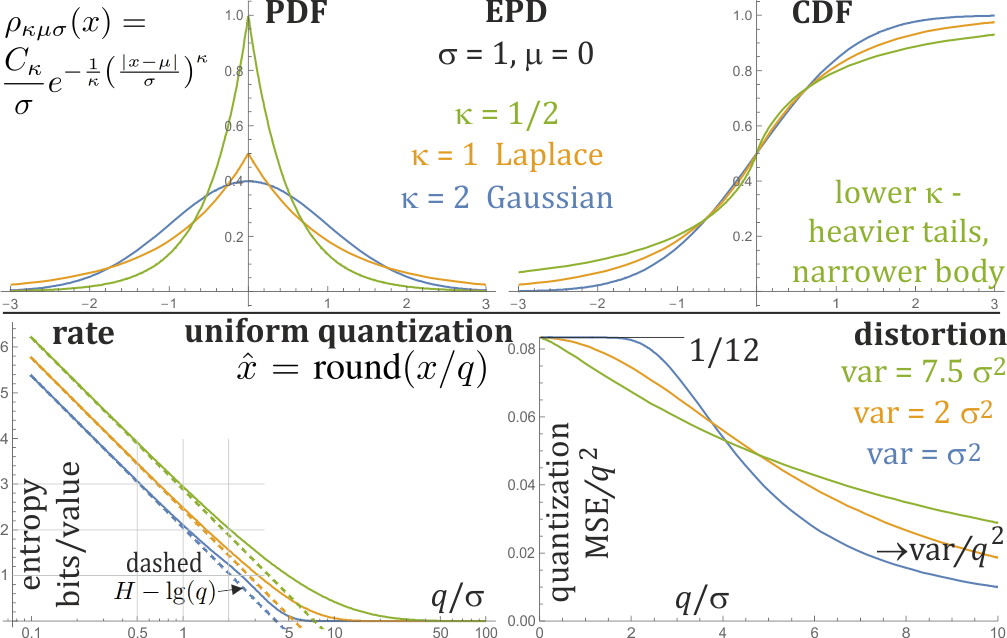}
        \caption{\textbf{Top}: probability distribution function (PDF) and cumulative distribution function (CDF) of exponential power distribution (EPD) for $\sigma=1,\mu=0$ and $\kappa=1/2,1,2$. \textbf{Bottom}: its rate and distortion for uniform quantization  $\hat{x}=\textrm{round}(x/q)$: for low $q/\sigma$ entropy is $\approx H-\lg(q)$ bits/value (dashed), distortion MSE $\approx q^2/12$. However, for large $q$ these formulas would exceed the boundaries: entropy cannot get below 0, MSE is bounded from above by variance. Therefore, for large $q$ (above $\sim \sigma$) behavior of these parameters is deformed not to exceed the boundaries, but to approach them asymptotically instead.}
       \label{EPDdist}
\end{figure}

While finding analytic MSE formula for general $\kappa$ seems difficult, Figure \ref{EPDdist} presents numerically found behavior: for small $q$ we have $MSE\approx q^2/12$, which is reduced especially when $q$ exceed $\sigma$, in a bit different way for various $\kappa$, asymptotically approaching variance.\\

\begin{figure}[t!]
    \centering
        \includegraphics{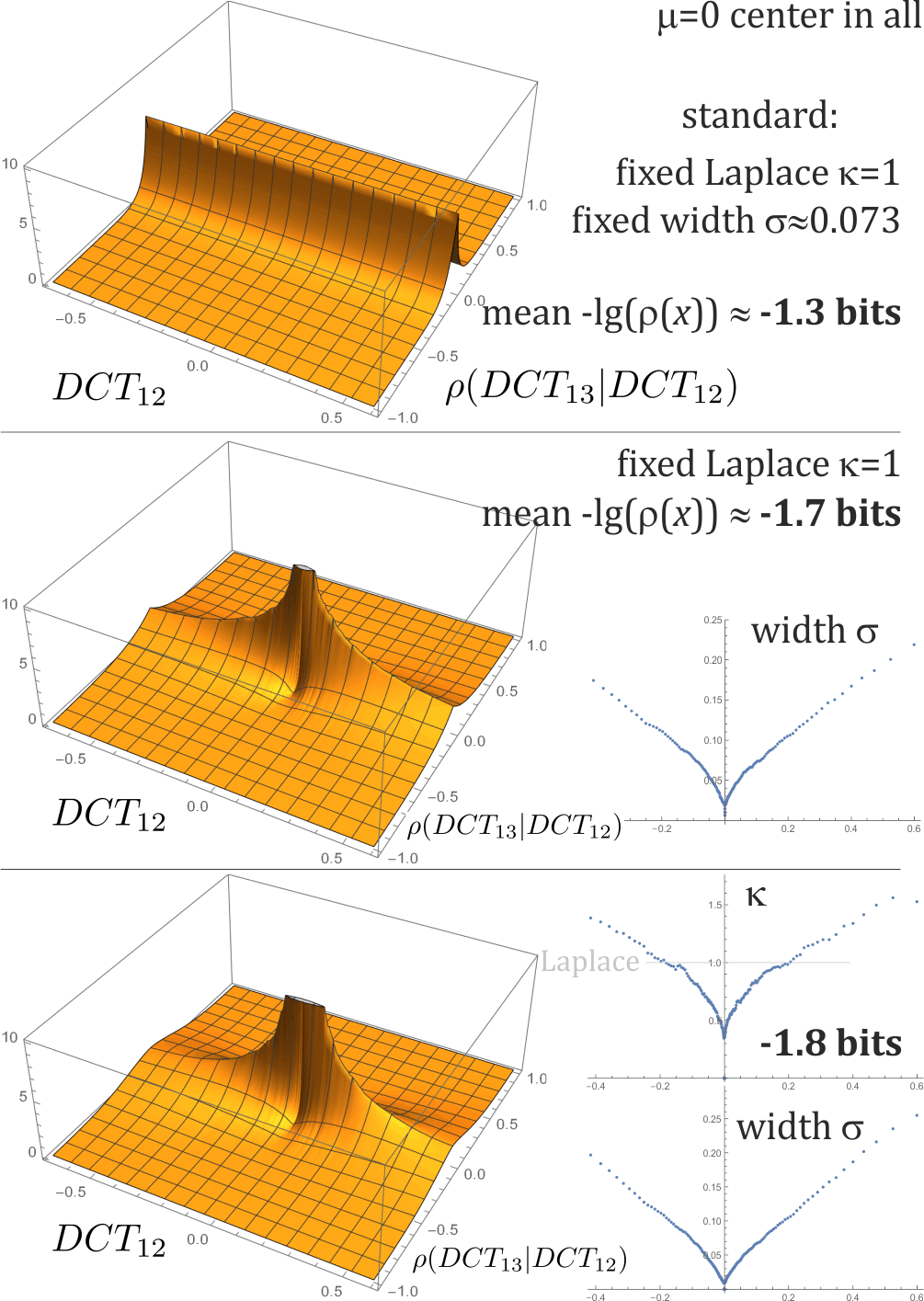}
        \caption{While DCT has nearly removed correlations between coefficients, there have remained higher statistical dependencies like homoscedasticity - exploiting of which can bring essential savings for data compression. Here is example of choosing probability distribution for $DCT_{13}$ coefficient, if knowing (already decoded) value of $DCT_{12}$. The plots were calculated by sorting $(DCT_{12},DCT_{13})$ pairs over the first coordinate and focusing on overlapping size 5000 ranges of pairs - mean first coordinate over such range is treated as $DCT_{12}$ in the plots. For second coordinate, in each range there is ML estimated parametric distribution. \textbf{Top}: standard assumption that distribution of $DCT_{13}$ is from Laplace distribution independent of $DCT_{12}$ value. \textbf{Middle}: for each range there is independently estimated Laplace distribution, we can see that the larger $|DCT_{12}|$, the larger width $\sigma$ should we choose, with nearly linear dependence. This way we get $\approx 0.4$ bits/value savings. \textbf{Bottom}: analogously, but estimating more general EPD instead, we can see that additionally $\kappa$ should grow with $|DCT_{12}|$, increasing savings to  $\approx 0.5$ bits/value. However, trials to essentially improve with varying $\kappa$ were unsuccessful so far (also much more costly), hence there is only used middle $\kappa=1$ case.}
       \label{cond}
\end{figure}

\begin{figure}[t!]
    \centering
        \includegraphics{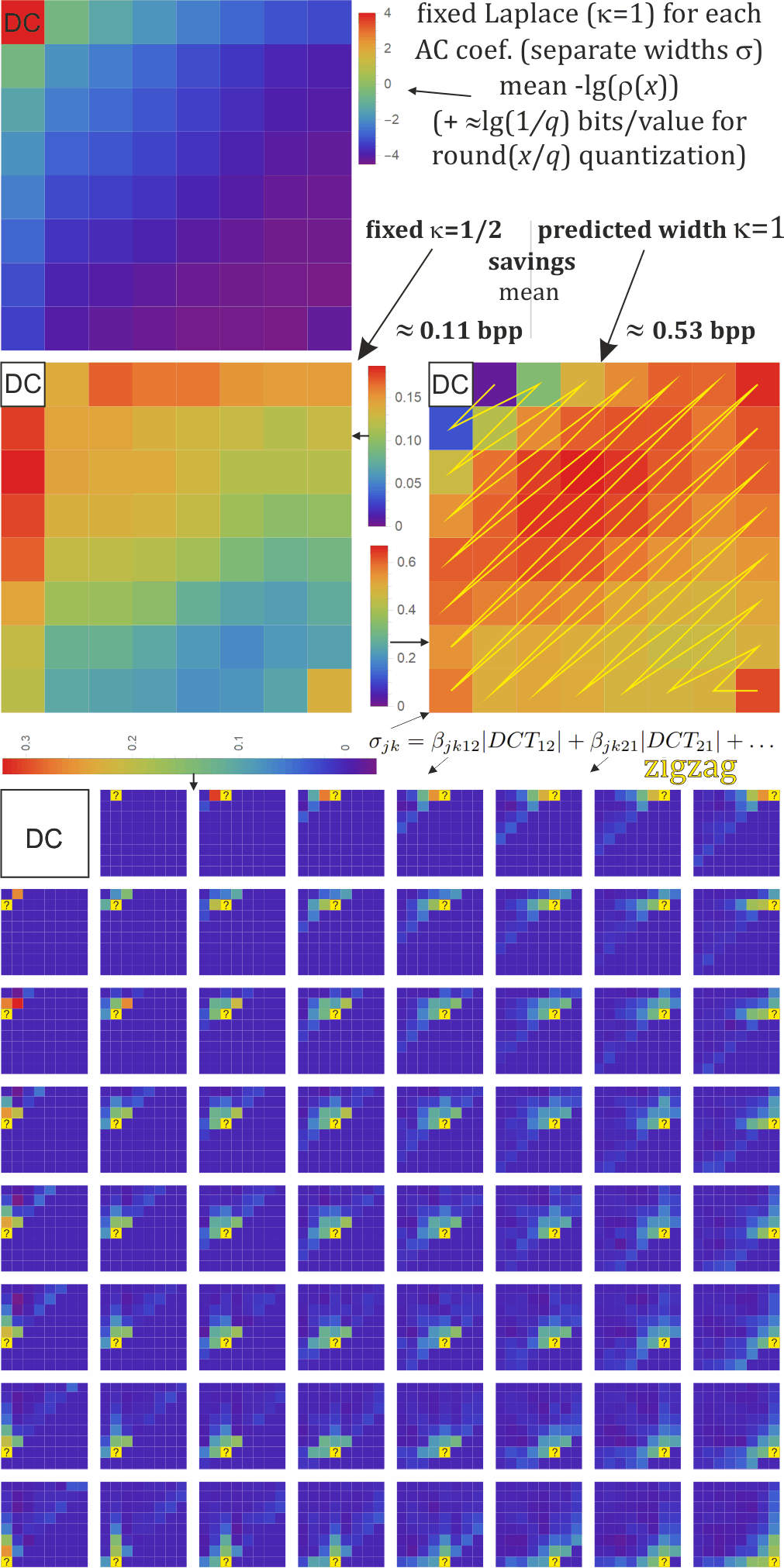}
        \caption{\textbf{Top}: minus log-likelihood (mean $-\lg(\rho(x))$) for AC coefficients, each independently estimated with Laplace distribution. For quantization as $\textrm{round}(x/q)$ we should add $\lg(1/q)$ to get approximate number of required bits/value. \textbf{Middle} left: discussed savings from using EPD $\kappa=1/2$ instead, on average $\approx 0.11$ bits/pixel (sum divided by 64 pixels). Middle right: savings from choosing width $\sigma$ based on already decoded coefficients, leading to much larger $\approx 0.53$ bits/pixel mean savings for using fixed $\kappa=1$. \textbf{Bottom}: visualized coefficients for this prediction of width $\sigma$, obtained from linear regression to minimize mean square error of absolute value of predicted coefficient $|DCT_{jk}|$.}
       \label{dctzz}
\end{figure}

While ML estimation of $\kappa$ is more difficult, in practice we can often use it as constant - optimized for a given situation, like general AC coefficients, or maybe 63 individual ones for each AC coefficient of 8x8 DCT as considered in Fig. \ref{epd}. In many cases like AC coefficients here we can also assume $\mu=0$, alternatively there can be used MLE approximation as mean value (exact for $\kappa=2$ Gaussian distribution), or median value (exact for $\kappa=1$ Laplace distribution), we can also predict it from a context as discussed e.g. in ~\cite{param,param1}.

There remains the main estimation - of width parameter $\sigma$, what turns out quite simple:
\be \sigma^\kappa = \textrm{mean }|x-\mu|^\kappa \ee
which can be seen as generalization of the Laplace and Gaussian case, can be naturally turned into context-dependent~\cite{param} (e.g. in the next section) or adaptive~\cite{exp1} estimation for nonstationarity.

Here for all 8x8 DCT-II coefficients from 48 grayscale 512x512 images there was calculated log-likelihood for various $\kappa$ - results are presented in Fig. \ref{epd}. We can see that $\kappa=1/2$ fits AC data much better than standard $\kappa=1$, getting $\approx 0.1$ bits/value reduction. In contrast, DC coefficients have completely different behavior and treatment, here getting optimal $\kappa\approx 2.2$.

\section{In-block conditional distributions}
DCT transform decorrelates data in block e.g. $8\times 8$, making additional linear predictions between coefficients inside block rather impractical (experiments suggest $<0.01$ bpp savings). It still leaves opportunities for between-block predictions, for example exploiting assumption that DCT in $16\times 16$ block should also decorrelate well, what allows to predict values in its one of four $8\times 8$ subblocks based on already decoded three remaining $8\times 8$ subblocks. Some between-block prediction is explored in the next Section, here we focus on additional opportunities inside a single block for AC coefficients.

While DCT allows to exploit correlation of coefficients inside a block, we have also higher statistical dependencies, e.g. between widths of neighboring distributions like homoscedasticity in ARCH-like models. Turns out its exploitation for AC can lead to essential savings: $\approx 0.53$ bits/pixel (total savings for grayscale divided by 64 pixel block, for RGB might be up 3x larger).

Figure \ref{cond} shows estimated example of such statistical dependence - with nearly zero correlation, but high width dependence, also of $\kappa$ shape parameter if using EPD.  It leads to $0.4-0.5$ bits/value savings - by estimating width from single already decoded coefficient. Figure \ref{dctzz} shows example of savings and coefficients from automated estimation from already decoded coefficients.

Specifically, for presented standard zigzag order, for each AC $j,k$ position we calculate width of Laplace distribution $\sigma_{jk}$ (centered in $\mu=0$) as linear combination of absolute values of previous (already decoded) coefficients in zigzag order:
\be \sigma_{jk}=\beta_{jk0}+\beta_{jk12} |DCT_{12}|+\beta_{jk21} |DCT_{21}|+\ldots\label{sig}\ee
Where as discussed in \cite{param,param1}, $\beta$ coefficients are found with the least squares linear regression to minimize distance to absolute value of predicted coefficient:
$$\arg\min_\beta \textrm{mean } \left(|DCT_{jk}|-\beta_{jk0}-\beta_{jk12} |DCT_{12}|-\ldots\right)^2$$
Surprisingly, the intercept term $\beta_{jk0}$ has nearly negligible effect - we can well estimate $\sigma$ from previous coefficients alone. Using DC coefficient $DCT_{11}$ for this prediction gives no improvement, however, e.g. gradients of DC coefficients of neighboring blocks can be useful - in the next Section we will get improvement by using absolute values of residues: errors from predictions $|x-\mu|$, instead of actual values.

We should avoid negative $\beta$ coefficients as they could lead to problematic negative $\sigma$. Directly applying such $\sigma$ estimation would require a few dozens of multiplications per value - in practice there are needed approximations reducing it to a few, for example using only a few neighboring already decoded values, maybe also some hidden states like $\sigma$ estimators found for these neighbors (this way containing combination of all previous, can be also states representing already decoded neighboring blocks), using only positive $\beta$ coefficients.

As discussed in \cite{param}, we can also apply optimized powers to terms in (\ref{sig}), e.g. for EPD directly estimating $\sigma^\kappa$ instead like variance $\sigma^2$ for Gaussian $\kappa=2$. There were performed some initial trials, also of predicting $\kappa$ as in bottom of Fig. \ref{cond}, but without getting essential improvements.

\section{Predictions between DCT blocks}
Not wanting blocking artifacts for block boundaries, there can be added constraints ensuring similar values for boundary pixels, also allowing to improve the compression ratio. Inexpensive way to realize it in practice are predictions for new block based on already decoded especially neighboring blocks.

\begin{figure}[t!]
    \centering
        \includegraphics{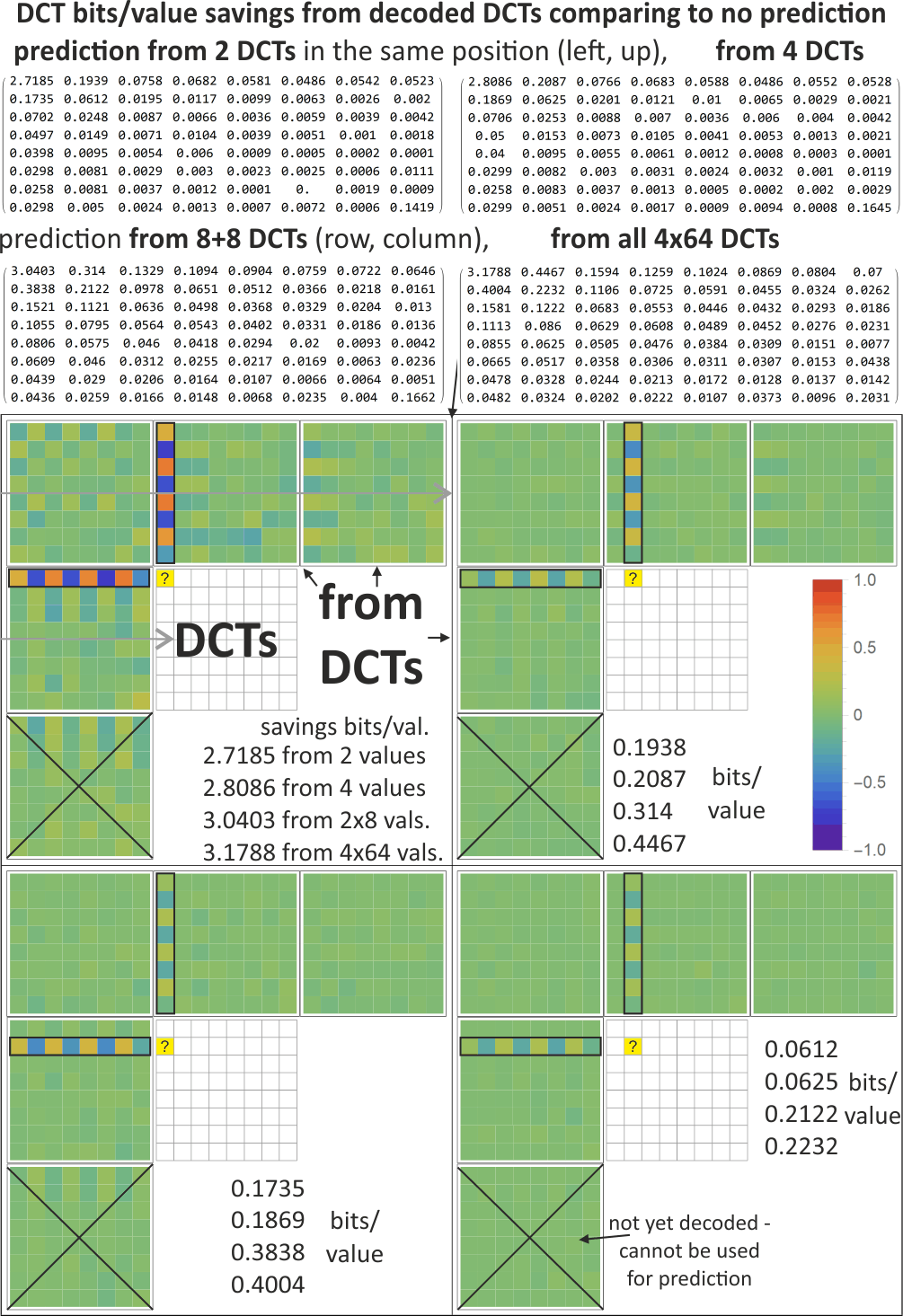}
        \caption{\textbf{Top}: DCT savings in bits/value as $\lg(\sqrt{MSE_0/MSE_{pred}})$ where $MSE_0$ is variance (mean-squared error from expected value) - no prediction, and $MSE_{pred}$ after various types of linear prediction from DCT coefficients of decoded neighboring blocks. Up-left: from 2 values in the same $jk$ position in block toward left and up. Up-right: analogously but using values from all 4 marked blocks (up, left, up-left, up-right). Down-left: using $8+8$ values from emphasized $j$-th row in left neighbor and $k$-th column in neighbor above, approximately corresponding to boundary values of these blocks as in Fig. \ref{prval}. Down-right: using entire model from $4\times 64$ values of 4 neighbors, presented (with addition of left-down) for 4 DCT coefficients in \textbf{bottom} of diagram (from linear regression). Mean savings (divided by 64) between the worst and best predictions here is $\approx 0.05$ bits/value, what is tiny comparing to $\approx 0.5$ from width prediction, however, in practice there are often encoded only for low frequencies - for which savings are quite large here. Additional advantage of high frequency predictions is reduction of blocking artifacts.     }
       \label{prdct}
\end{figure}

We would like to predict values of DCT coefficients (also their uncertainty/width $\sigma$), suggesting to directly use DCT coefficients of neighboring blocks as the context, preferably using linear combination for inexpensive calculation. Density plots in Figure \ref{prdct} show such found coefficients from least squares linear regression. We can see characteristic rows/columns with alternating coefficients - intuitively they correspond to decoding DCT coefficients into values for pixels adjacent to the new block.

Figure \ref{prdct} also contains bits/value savings from such predictions (translated from MSE improvement) - for DC this saving is huge, quickly weakening down to zero for higher frequencies - in practice we can focus only on prediction for low frequencies, however, doing it also for higher frequencies should reduce blocking artifacts. Four $8\times 8$ matrices in this Figure present savings from approaches of growing computational cost: from using single DCT values (of $jk$ position as the predicted one) in 2 (up, left) or 4 (up, left, up-left, up-right) neighboring blocks, then using  $8+8$ DCT coefficients only from the marked rows/columns of corresponding positions, and finally using all $4\times 64$ DCTs from these 4 blocks - which should be already decoded if scanning blocks in succeeding horizontal lines (gray). Down-left block is shown to better understand dependencies, but cannot be used for prediction.

\begin{figure}[t!]
    \centering
        \includegraphics{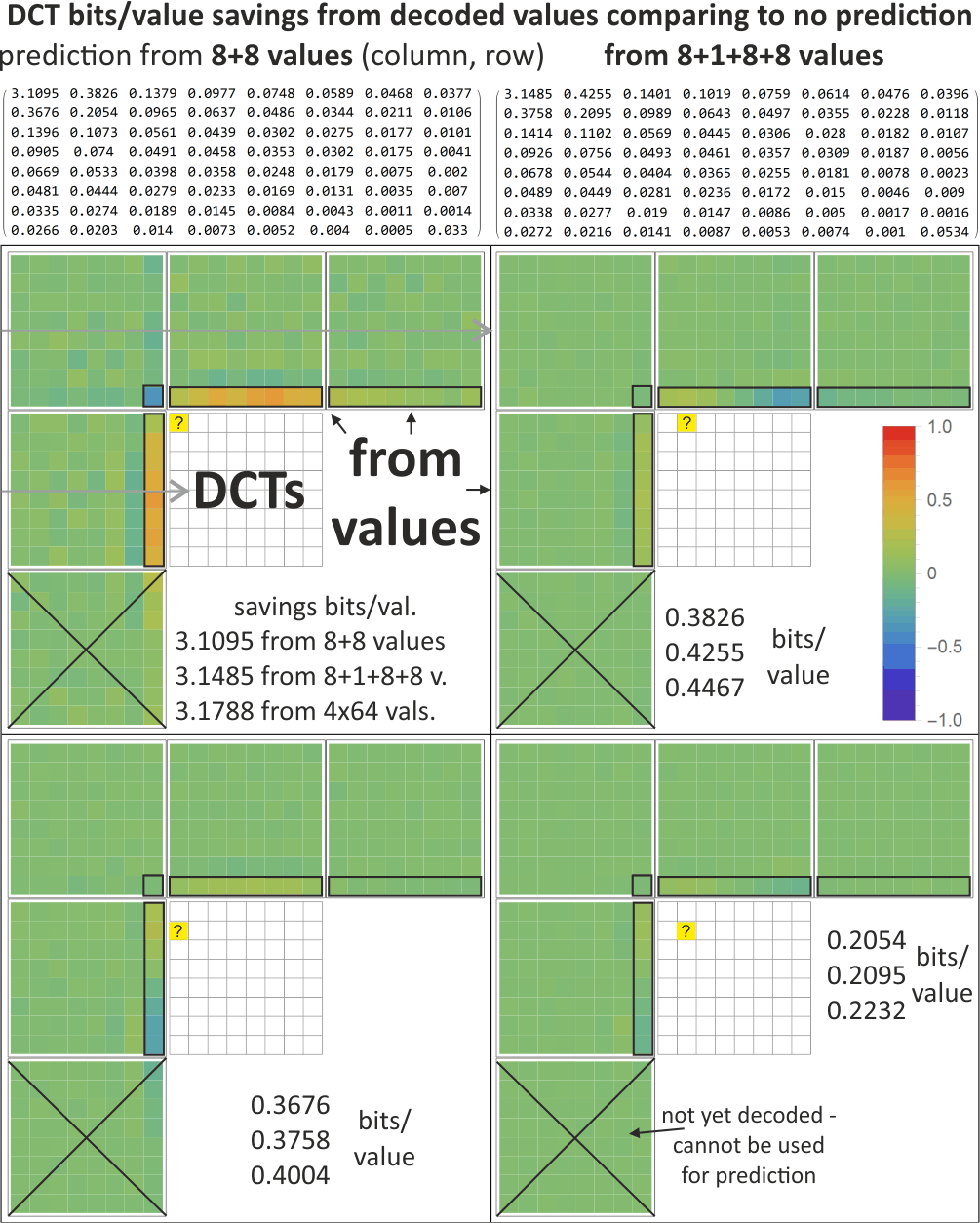}
        \caption{Analogously as Fig. \ref{prdct}, but predicting from decoded pixel values of neighboring blocks (instead of their DCTs). As we could expect, the most crucial here are the adjacent pixels: right columns in blocks toward left, bottom-right pixel in up-left block, and bottom row in blocks above. Alternating coefficients in Fig. \ref{prdct} approximately decode values of these boundary pixels. \textbf{Top} left: bits/value savings due to prediction from such $8+8$ values in blocks toward up and left. Top right: analogously using mentioned marked $8+1+8+8$ adjacent pixels from all 4 blocks. \textbf{Bottom}: linear coefficients using all $4\times 64$ values, leading to savings exactly as the best in Fig. \ref{prdct} (as DCT is a linear transform).}
       \label{prval}
\end{figure}
\begin{figure}[t!]
    \centering
        \includegraphics{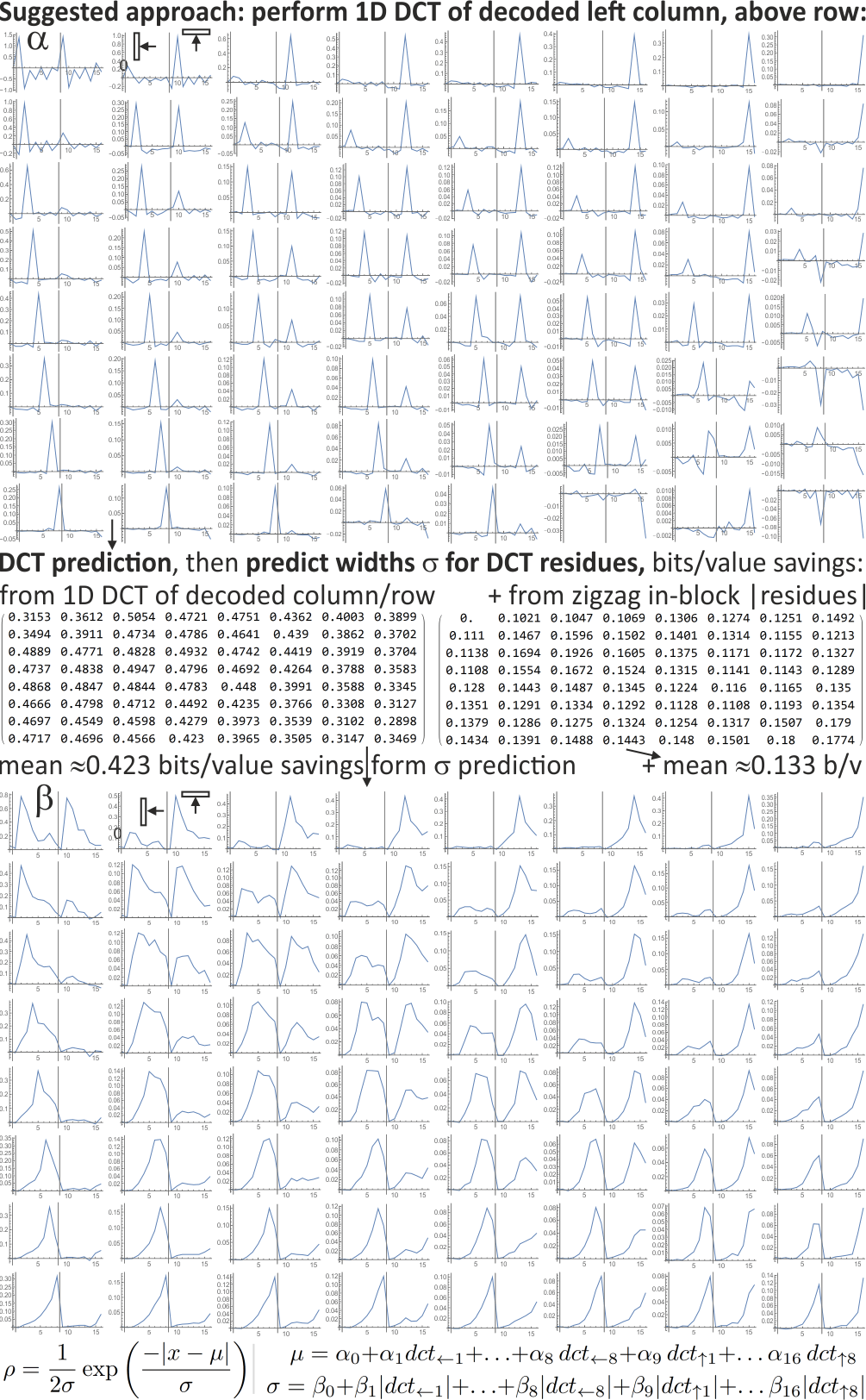}
        \caption{Suggested general approach for practical compromise between computational cost and compression level. Based on Fig. \ref{prdct}, \ref{prval}, it seems most effective to exploit values from adjacent pixels of already decoded blocks. Even more, e.g. these two figures suggest frequency dependence, so it seems valuable to calculate one-dimensional (1D) DCT (as $dct$) of such rows and columns (much less expensive than 2D DCT), and find linear predictions from them. \textbf{Top}: linear regression coefficients from predicting (2D) DCT coefficients of new block form 1D DCT coefficients of 8 adjacent pixels toward left, and of 8 adjacent pixels above. We can see domination of coefficients of corresponding frequencies - allowing to approximate such prediction with a few most significant terms. \textbf{Middle}: additional savings in bits/value from predicting width $\sigma$ of Laplace distribution for residues as linear combinations of absolute values of these 1D DCTs (denoted $dct$), with coefficients presented in \textbf{bottom} - we can again notice some frequency dependence, which allows for practical approximations. In Fig. \ref{ccasig} such models are reduced with CCA. Middle-right: additional savings in bits/pixel if also using absolute values of residues(!) of already decoded DCT coefficients in this block in zigzag order - as in Figure \ref{dctzz}. While it is quite costly, it might be worth to find some its practical approximation.  }
       \label{sigpred}
\end{figure}

To reduce computational cost, we should reduce context for such prediction e.g. from entire blocks to only adjacent pixels, focusing on them is also suggested by alternating coefficients in Fig. \ref{prdct}. Assuming that processed blocks are already decoded, we can use their final values instead of their DCTs. Figure \ref{prval} does it analogously, we can see that as expected, the adjacent pixels has turned out the most essential for such prediction. This figure also presents savings from using $8+8$ such values from left and up block, or from $8+1+8+8$ marked values from 4 decoded blocks - providing a bit more economical prediction than using DCTs of these blocks.

Coefficients in these marked columns and rows resemble Fourier coefficients, what suggests calculating one-dimensional DCT for right-most columns and lowest rows of decoded blocks to use for the prediction purposes, what computationally is much less expensive than standard 2D DCT. Using such 1D DCT coefficients denoted by $dct_{\leftarrow}, dct_{\uparrow}$ for the left and above neighbors, we can e.g. use model for parameters of used Laplace distribution for $DCT_{jk}$ as
$$\rho(x)\equiv \rho^{jk}(x) =\frac{1}{2\sigma} \exp\left(-\frac{|x-\mu|}{\sigma}\right)$$
\begin{small}
$$\mu= \alpha_0 +\alpha_1 dct_{\leftarrow 1}+\ldots+\alpha_8\, dct_{\leftarrow 8}+\alpha_9\, dct_{\uparrow 1}+\ldots \alpha_{16}\, dct_{\uparrow 8}$$
$$\sigma= \beta_0 +\beta_1 |dct_{\leftarrow 1}|+\ldots+\beta_8 |dct_{\leftarrow 8}|+\beta_9 |dct_{\uparrow 1}|+\ldots \beta_{16} |dct_{\uparrow 8}|$$
\end{small}
\noindent where coefficients are separate for each predicted $DCT_{jk}$ position (should be e.g. $\mu^{jk}, \alpha^{jk}, \sigma^{jk}, \beta^{jk}$) - they are presented in two $8\times 8$ arrays of plots in Fig. \ref{sigpred}. As previously, both were found by least-squares linear regression: $\mu$ directly from coefficients, $\sigma$ from absolute values of $dct$ to minimize MSE from absolute values of residues $|x-\mu|$. While they use all the $8+8$ values, corresponding frequencies are dominating - we can reduce used context to a few such dominating values to reduce computational cost.

Savings from such $\mu$ predictor alone are in top-left array in Fig. \ref{prval} (due to linearity, prediction from these values and their 1D DCT is the same). Additional prediction of $\sigma$ (comparing to use of fixed for each position) gives much larger additional mean savings: $\approx 0.423$ bits/value, presented in center-left of Fig. \ref{sigpred}. While quantization optimization might reduce it for high frequencies by mostly using 0 values, this uncertainty evaluation could still lead to relatively huge savings - provides opportunity often neglected in standard approaches.

Finally, center-left array in \ref{sigpred} shows additional savings if also using already decoded coefficients from this block in zigzag order as in Fig. \ref{dctzz} - while they are not useful for $\mu$ prediction (thanks to decorrelation property of DCT), they can help evaluating uncertainty/width $\sigma$: by using absolute values of residues(!) as additional context. Using residues $|DCT_{jk}-\mu^{jk}|$ instead of $|DCT_{jk}|$ gives a better prediction - evaluation of certainty of $\mu$ prediction.

There was not yet tested EPD optimization here. There also remains question of handling RGB colors, a first suggestion is expanding Fig. \ref{sigpred} approach with 1D DCT context from $2\times 8$ to $3\times 2 \times 8$ for all 3 colors (presented in Fig. \ref{fin}), then maybe approximation using only dominating coefficients - let us now discuss using CCA for this purpose.

\section{Canonical correlation analysis (CCA)}
Classical CCA~\cite{CCA} technique is a natural tool for practical approximations of discussed inexpensive linear models, optimizing bottleneck to reduce computational cost. Let us briefly introduce it.

For two multidimensional random variables $X$ and $Y$, in CCA we search for direction pairs $(a,b)$ maximizing correlation:
$$\operatorname*{argmax}_{a,b}\ \textrm{corr}(a^T X, b^T Y)$$
Applied multiple times, it would lead to a chosen size orthonormal set of vectors for $X$ and $Y$ - we can treat as features for prediction.

In practice it is calculated by whitening the variables - multiplication by ($C^{-1/2}$) matrix to get normalized variables of unitary covariance matrix, then perform SVD (singular value decomposition) of cross-covariance matrix for such normalized variables.

Specifically, for $\mu_X=E[X], \mu_Y=E[Y]$ expected values vectors, we need covariance matrices:
\begin{small}
$$C_{XX}=E[(X-\mu_X)(X-\mu_X)^T],\quad C_{YY}=E[(Y-\mu_Y)(Y-\mu_Y)^T]$$
\end{small}
$$C_{XY}=E[(X-\mu_X)(Y-\mu_Y)^T]\qquad\quad C_{YX}=C_{XY}^T $$
Performing SVD for cross-covariance matrix of whitened variables, and returning to the original variables, we get
$$a\textrm{ is an eigenvector of }C_{XX}^{-1} C_{XY} C_{YY}^{-1} C_{YX} $$
\be b\textrm{ is proportional to } C_{YY}^{-1} C_{YX} a\ee
In practice we use some number of such vector pairs corresponding to the highest eigenvalues: the strongest dependencies.
\begin{figure}[t!]
    \centering
        \includegraphics{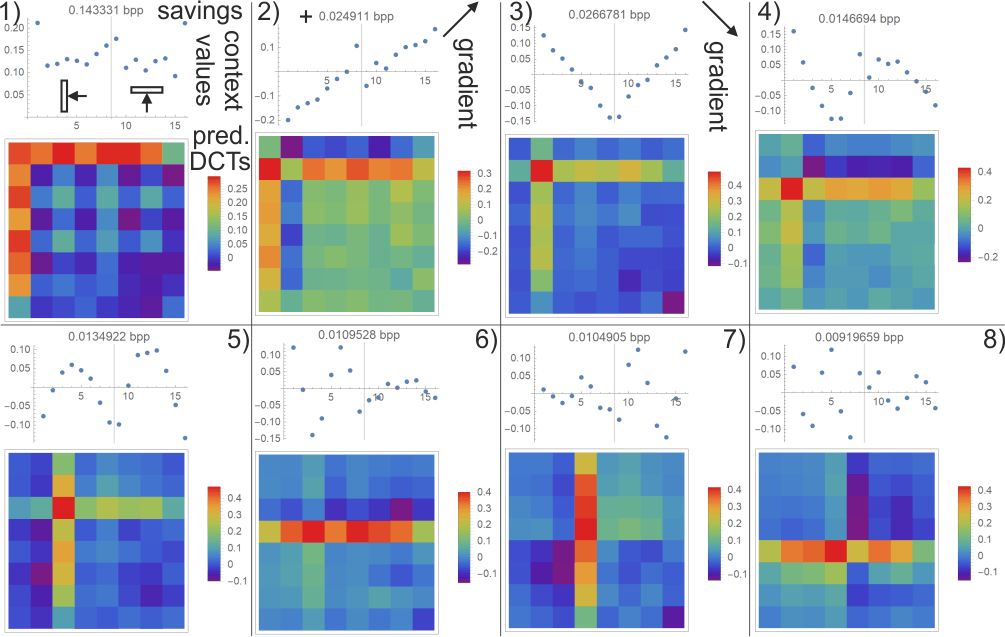}
        \caption{Practical approximation for value prediction using CCA - automatically optimized inexpensive  bottleneck for linear prediction. It uses context as values of pixels in column toward left and row above: the features are linear combinations presented in plots above, with DCT prediction contributions presented in $8\times 8$ maps below. The savings are in bits/pixels for RGB if adding given feature, e.g. 2) gets $\approx 0.143+0.025$ bpp savings. The 1) is nearly mean over the context - it corresponds to prediction of DC and ACs corresponding to functions constant in one direction (red/orange). Then 2), 3) correspond to two types of gradients. Further eigenvectors automatically got dependence from higher frequencies. }
       \label{cca}
\end{figure}

Figure \ref{cca} contains example of value prediction using such analysis (instead of 1D DCTs in Figure \ref{sigpred}). Using only 1) feature from the context: nearly its mean, adding corrections to DCTs as in the color map, we get $\approx 0.14$ bpp savings. Adding 2), 3) features corresponding to different types of gradients, we get additional $\approx 0.05$ bpp. Adding the following 5 features corresponding to higher frequencies, we get another $\approx 0.06$ bpp. Here we assume that all DCTs are modified with such prediction, in practice we can approximate by adding only to a few low frequency ones.

\begin{figure}[t!]
    \centering
        \includegraphics{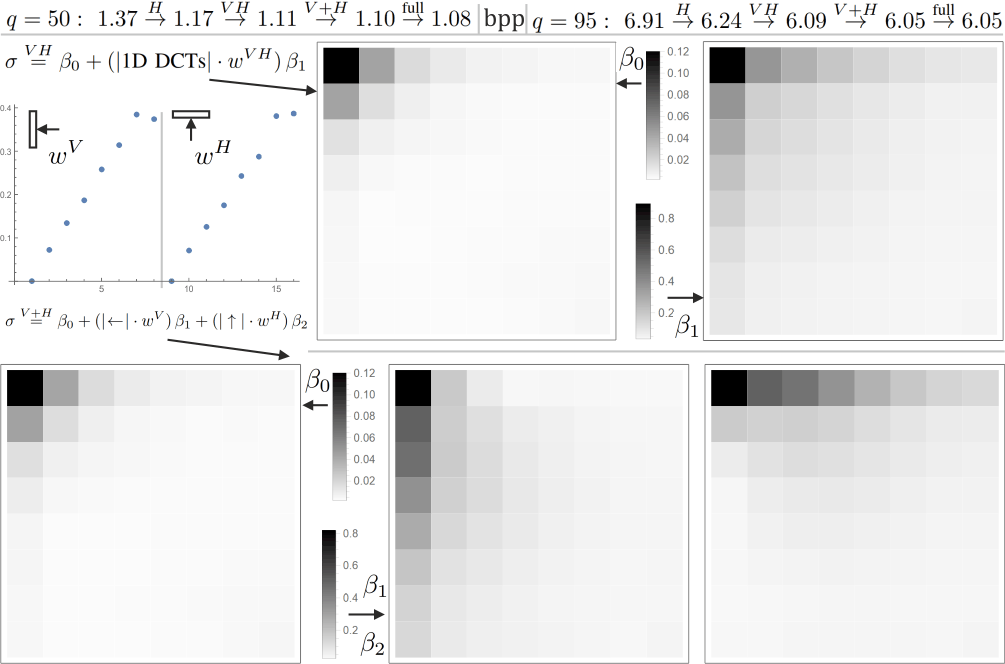}
        \caption{CCA for width $\sigma$ prediction: 1-2 feature replacement for full "predict widths from 8+8" in Fig. \ref{fin} (which uses huge models as in Fig. \ref{sigpred}). Top left: weights $w$ from first CCA eigenvector for (concatenated) absolute values of 1D DCT of column to left ($V$), and of row above ($H$), we can see these weights have nearly linear frequency dependence, what allows to generalize to larger, also rectangular blocks. Such basic $VH$ width prediction adds $\beta_0$ to $\beta_1$ times such single feature evaluating local noise level. Both $\beta$ depend on frequencies - there are presented their maps for various DCTs (in practice they can be divided by corresponding quantization coefficients). Bottom: two feature model with separate vertical and horizontal noise level evaluation, by separately multiplying vertical and horizontal $|$1D DCTs$|$. At the top there are written mean bits/pixel between two levels of Fig. \ref{fin}: 1.37 or 6.91 (for quality 50 or 95) without sigma prediction, 1.08 or 6.05 for costly full model. We can see that such cheap 1 or 2 feature approximations can get very close to the full model. There is also written evaluation for $H$ model using only the row above, what is convenient for parallelization, but gives slightly worse compression. }
       \label{ccasig}
\end{figure}

It is tempting to use CCA also for width $\sigma$ prediction. It can be thought of as evaluation of local noise level, should use only positive weights e.g. to ensure $\sigma > 0$. Due to orthogonality, only the first CCA eigenvector can have only positive weights. However, as we can see in Fig. \ref{ccasig}, fortunately such single noise level feature already leads to nearly the same evaluation as the previously discussed full model using 1D DCTs of already decoded neighboring pixels. It can be split e.g. into horizontal and vertical noise evaluation, getting two features allowing for slight improvement. For parallelization it is convenient to use only the horizontal part from the previous row, what leads to intermediate compression ratios.



\section{Adding uniform quantization}
\begin{figure}[t!]
    \centering
        \includegraphics{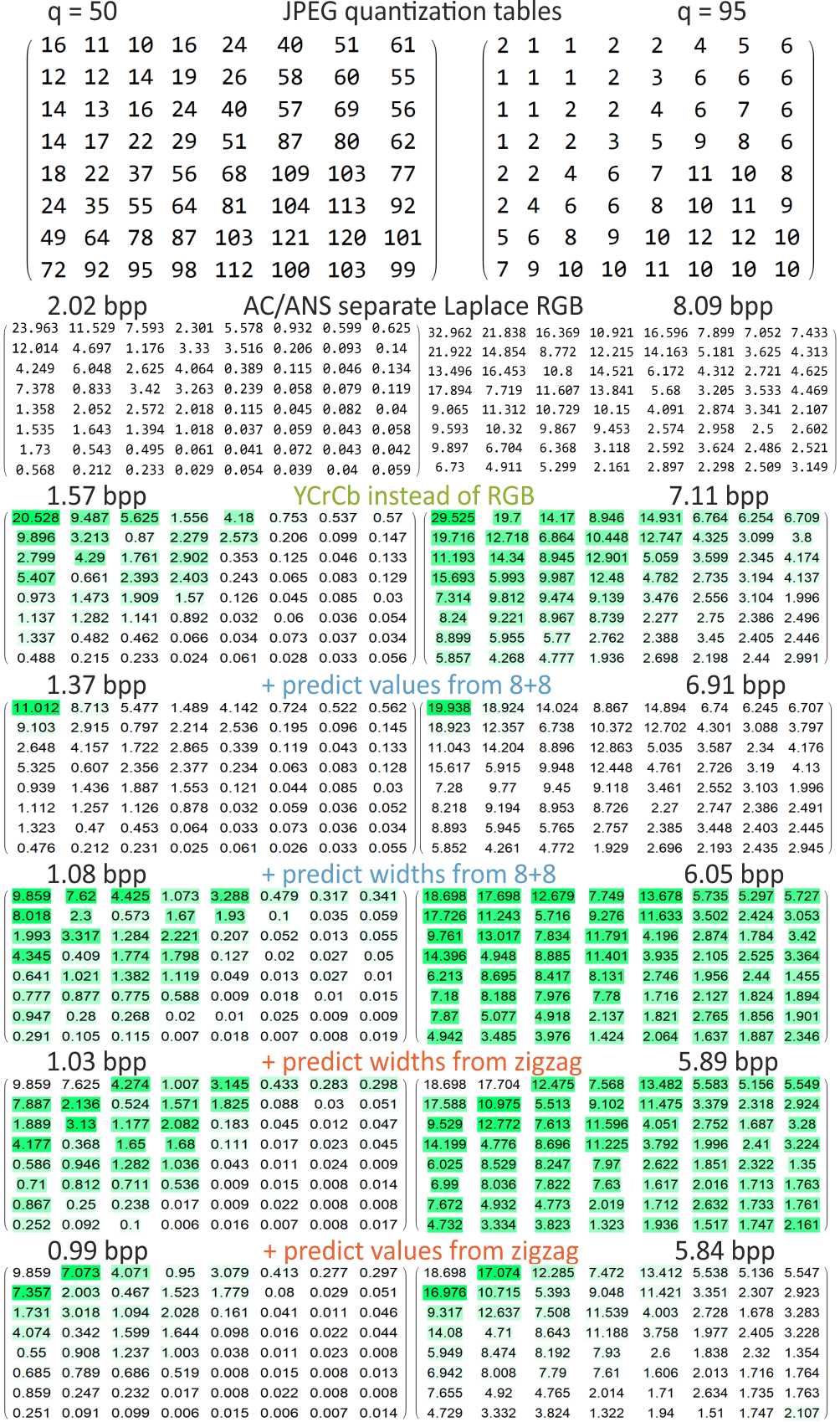}
        \caption{Evaluation for DCT encoding using two JPEG~\cite{jpeg} quantization tables $\hat{x}_{cij}=\textrm{round}(256\, x_{cij} / Q_{ij})$ on dataset: all channels $c$ are treated the same, each value is entropy coded (AC/ANS) assuming Laplace distribution chosen separately for each $(c,i,j)$ (a bit differen than JPEG: which uses luma and chroma quantization tables, usually prefix codes, marking zigzag position to further use all zeros). From the top, the first directly encodes RGB channels, matrices contain numbers of bits summed over 3 channels, their sums divided by 64 is mean bits/pixel. Second: the same but transforming to less correlated YCrCb channels first. Third additionally uses linear prediction from the same channel of decoded neighboring column toward left, and row above. Fourth additionally uses linear prediction of widths from absolute values of 1D DCT coefficients of such column and row. Fifth additionally uses absolute values of already decoded  DCT coefficients in this block in zigzag order as context for predictions of widths. Sixths analogously also predicts values from already decoded DCT. The above use only one corresponding channel for predictions, improvement from using all 3 is negligible. The last two improvements are computationally costly, but their practical approximations should give similar savings. Green color visualizes  bits/value improvement distribution (comparing to one method up) for various frequencies - while value prediction mainly improves low frequency coefficients (their exploitation should have higher priority), width prediction improves nearly all.}
       \label{fin}
\end{figure}
There remains difficult question of adding quantization, preferably chosen with rate-distortion optimization, and flexible to inexpensively vary it depending on region of image.

Basically there are three levels of evaluations for predicting encoded value $X$ based on context $Y$, the first two are approximate but useful for various stages of modelling:
\begin{enumerate}
  \item For predicting value alone as $X\approx \mu(Y)$ estimator: $\lg(\sqrt{E[(X-E[X]]^2)/E[(X-\mu(Y))^2]})$ gives approximate for saved bits based on reduced MSE: comparing to context-free prediction as expected value $E[X]$.
  \item For parametric probability distribution $\rho_Y$: beside predicted center $\mu(Y)$ above, also e.g. scale parameter $\sigma(Y)$ of Laplace distribution, maybe also other parameters like $\kappa$ for EPD (fixed or context dependent), leading to log-likelihood evaluation: $-E[\lg(\rho_Y(X))]$ which after adding $\lg(1/q)$ gives approximate number of bits/value for step $q$ uniform quantization.
  \item For discretized distribution: the above uniform quantization approximation is convenient for flexible rate-distortion optimization, but is appropriate only for small $q/\sigma$ ($\sim$ below 1) like for low frequency DCT coefficients. Generally we need entropy of final discrete distribution, for Laplace there are analytical formulas (Section 2, plotted in Fig. \ref{EPDdist}) - it is crucial e.g. for high frequency DCT coefficients often quantized to 0 value, exploiting the fact that MSE is bounded by variance.
\end{enumerate}
So far we have focused on the first two: first if only predicting value, second if entire parametric distributions - additionally width $\sigma$ for used Laplace distribution. The last one requires choosing quantization tables for DCT, preferably with flexible rate-distortion, what is generally a difficult problem, discussed e.g. in \cite{jq}.

For simplicity, to get overview of possible optimizations of techniques discussed here, in Fig. \ref{fin} there are presented required bits/pixel for using standard JPEG quantization tables $Q$ for two quality levels: 50 and 95. The approach here is simplified comparing to JPEG: each (position,channel) pair is separately modelled, quantized the same way for each channel (JPEG uses different luma, chroma tables), then we assume AC/ANS entropy coding for each value (JPEG usually uses prefix codes).


\subsection{Practical entropy coding for $(\mu,\sigma)$ prediction}
For building data compressors, let us  focus here on the basic case: prediction of value $\mu$ and scale parameter $\sigma$, for example based on the context (or adaptive), using Laplace distribution and step $q$ uniform quantization.

It is natural to subtract the predicted value first: encode residue $x-\mu$ assuming distribution centered in zero.

For residue we know both $\sigma$ width, and quantization step $q$. To use 1-parameter family of coding tables, we can rescale horizontal axis of such distribution density: use $Q=q/\sigma$ normalized quantization step for width 1 Laplace distribution centered in 0. Its CDF is:
$$\textrm{CDF}(v) = \left(\textrm{sign}(v) (1-\exp(-|v|))+1\right)/2$$
Both encoder and decoder need to prepare AC/ANS tables for probability distributions of quantized values $\{(i,\textrm{Pr}(i)):i\in \mathbb{Z}\}$, in practice bounded to a finite sets: grouping tails into single symbols. We need such distributions for some fixed discrete optimized set of $Q$, e.g. $Q=0.1,0.2,0.3,...$ parameters:
$$\textrm{Pr}(i) = \textrm{CDF}((i+0.5)Q) - \textrm{CDF}((i-0.5)Q)$$
Based on the context, we choose entropy coding table as discretized $Q=q/\sigma$, e.g. table number $\textrm{round}(10 Q)$
$$\textrm{round}((x-\mu)/q)\textrm{ entropy coded with table for } Q=q/\sigma$$

Alternative and probably more convenient view is analogously working with inverted $\Sigma:=Q^{-1}=\sigma/q$ determining the coding table, e.g. to use $\textrm{round}(10 \Sigma)$-th coding table for given coefficient. It allows to imagine varying $\sigma$ width of Laplace distribution, while quantization step is fixed: to 1 for $\Sigma$. This way we can find and use statistical models directly for quantized coefficients $(\textrm{round}(x/q))$ as integer numbers: estimate $\sigma$ for them, and treat it as $\Sigma$ (includes division by $q$): use e.g. $\textrm{round}(10 \sigma)$-th coding table. Alternatively, especially for varying quantization tables, we can also use model for $\sigma$ instead (as discussed throughout this article), then divide it by $q$ to get $\Sigma=\sigma/q$, what for linear models can be optimized by dividing their linear coefficients, e.g. $\sigma = \beta_0 + \beta_1 |c|$  replaced with $\Sigma = (\beta_0 / q) + (\beta_1 / q) |c| $.

We can analogously use other parametric distributions like EPD, for which CDF formula is in (\ref{EPDrho}), $\kappa$ shape parameter can be chosen or predicted, in the latter case requiring 2 parameter distribution family: both for $\Sigma=\sigma/q$ (or $Q=1/\Sigma$) and $\kappa$ e.g. $\kappa=0.5,0.6,\ldots,1.5$.

\subsection{Sigma quantization - preferably nonuniform}
We need to prepare entropy coding tables for some discrete set of $\Sigma=\sigma/q$, bounding error of such approximation e.g. to 0.01 bits/pixel, hence $E=1/300$ bits/value for RGB. Then we can e.g. use modified James Bonfield rANS implementation\footnote{\url{https://github.com/jkbonfield/rans_static}} for order 1 Markov model with prepared (value, number of coding table) sequence for the entire e.g. 8x8 DCT block.

\begin{figure}[b!]
    \centering
        \includegraphics{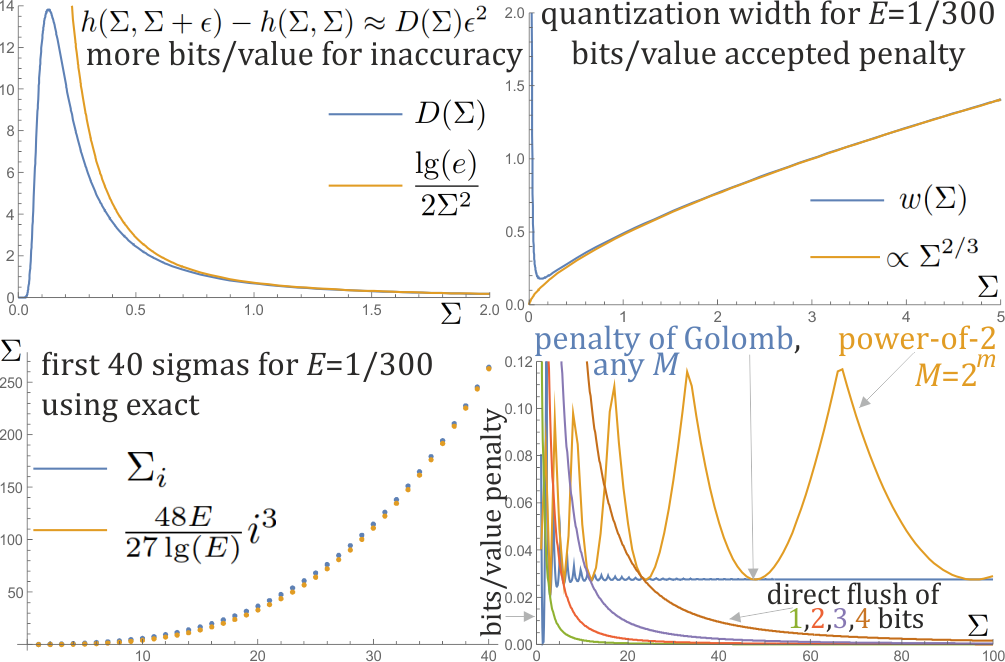}
        \caption{Top-left: inaccuracy penalty for $\Sigma$ approximation. Top-right: widths for quantization ranges for chosen penalty level. Bottom-left: some found $\{\Sigma_i\}$ quantization for coding tables. Bottom-right: bits/value penalties for using general Golomb coding, $M=2^m$ power-of-2 Golomb coding, and flushing $m=1,2,3,4$ least significant bits.}
       \label{sigma}
\end{figure}

We have width $\Sigma$ Laplace distribution centered in 0, uniformly quantized with step 1, getting two-sided geometric distribution of quantized values $(p_x = \int_{x-1/2}^{x+1/2} \rho(y)\, dy)$:
$$p_0^\Sigma =1-e^{-\frac{1}{2\Sigma}}\qquad x\in \mathbb{Z}\backslash \{0\}:\ p_x^\Sigma = e^{-\frac{|x|}{\Sigma}} \sinh\left(\frac{1}{2\Sigma}\right)$$
It allows to analytically find cross entropy - mean used bits/value for entropy coding assuming $\Sigma_q$ distribution, for sequence from $\Sigma_p$ distribution:
$$h(\Sigma_p,\Sigma_q)=\sum_{x\in\mathbb{Z}} p_x^{\Sigma_p} \lg(1/p_x^{\Sigma_q})\ \textrm{bits, leading to penalty:}$$
$$h(\Sigma,\Sigma+\epsilon)-h(\Sigma,\Sigma) = D(\Sigma)\, \epsilon^2 +O(\epsilon^3)\ \textrm{bits for:}$$
$$D(\Sigma)=\left(\frac{3e^{1/(2\Sigma)}+e^{1/\Sigma}+e^{3/(2\Sigma)}-1}{8\,(e^{1/\Sigma}-1)^2\, \Sigma^4}\right)\lg(e)\to \frac{\lg(e)}{2\Sigma^2}$$
This formula and its $\propto \Sigma^{-2}$ approximation are plotted in top-left of Fig. \ref{sigma}. Asymptotic $\propto \Sigma^{-2}$ behavior can be easily calculated from continuous Laplace distribution - the difference comes from quantization, is essential for $\Sigma\in (0,1)$ and practically negligible for $\Sigma > 1$. 

Using $h(\Sigma,\Sigma+\epsilon)-h(\Sigma,\Sigma)\approx D(\Sigma) \epsilon^2$ approximation, mean penalty in $[\Sigma-w/2, \Sigma+w/2]$ width $w$ range around chosen $\Sigma$, assuming uniform $\Sigma$ distribution inside is:
$$E\approx D(\Sigma) \int_{\Sigma-w/2}^{\Sigma+w/2} \epsilon^2 d\epsilon = \frac{1}{24} w^3\, D(\Sigma)\quad \textrm{bits/value}$$
$$\textrm{getting}\quad w(\Sigma) = \sqrt[3]{\frac{24 E}{D(\Sigma)}} \to \sqrt[3]{\frac{48 E}{\lg(e)}}\, \Sigma^{2/3}$$
as approximate width of sigma quantization around $\Sigma$ for $E$ bits/value accepted penalty level. Choosing $E=1/300$ bits/value corresponding to RGB 0.01 bits/pixel, we get top-right plot in Fig. \ref{sigma} for $w(\Sigma)$ and $\propto \Sigma^{2/3}$ approximation.

We can use it to optimize sigma quantization for chosen $E$ accepted penalty level. For example start with $\Sigma_1 \approx 0.1$ and then use $\Sigma_{i+1} =\Sigma_i + w(\Sigma_i)$ recurrence. For its approximation we can use $w(\Sigma)\propto \Sigma^{2/3}$ and solve $\frac{\partial}{\partial i}\Sigma=w(\Sigma)$ differential equation, getting $\Sigma_i \approx \frac{48 E}{27 \lg(E)} i^3$ simple behavior. Both result of recurrence and approximated are presented in bottom-left of Fig. \ref{sigma}, showing good agreement of approximation, but requiring caution for small $\Sigma<1$.

The largest $\Sigma$ are for DC coefficients, in sigma prediction reaching $\sim 200$ for 8bit values. For HDR much larger: $4\times$ for 10bit, $16\times$ for 12bit and so on. While it is tempting to just use uniform sigma quantization, nonuniform should give better compression, e.g. $i=\textrm{round}(c \sqrt[3]{\Sigma})$-th coding table for some constant $c$: by directly using cube root, or putting behavior into table e.g. $i=\textrm{Sigma}[\textrm{round}(8 \Sigma)]$-th table.

\subsection{Entropy coding approximations: Golomb and direct flush}
We need to prepare entropy coding e.g. rANS tables for the chosen set of sigmas. However, even for very small $\Sigma$, the value could get quite large - might require to handle exceptions. For large $\Sigma$, especially for DC in HDR, the alphabet could become huge.

Standard e.g. rANS implementations use e.g. 256 size alphabet, what could be increased at cost of larger tables and slower processing speed. We could also use rANS twice instead, e.g. for $\textrm{mod}(x,2^m)$ and $\lfloor x/2^m \rfloor$ preparing two separate (geometric distribution) tables for these sigmas, each restricted e.g. to 256 size.

Approximations could reduce this cost, e.g.:
\subsubsection{Golomb code} For fixed parameter $M$, encode $\lfloor x/M \rfloor$ with unary code, and $\textrm{mod}(x,M)$ directly. It requires $1+\lfloor x/M \rfloor + \lg(M)$ bits, optimal for $\textrm{Pr}(x) \sim 2^{-x/M}$.

For quantized Laplace distribution, we could e.g. use rANS for some central values, and put the tail behavior in the two extreme values: denoting to decode distance from this value using Golomb code of $M$ parameter e.g. determined by number of coding table, this way also handling exceptions.

This tail of quantized Laplace distribution has geometric distribution: $\textrm{Pr}(x) = a^x\,(1-a)$ for $a=e^{-1/\Sigma}$. It suggests to use $M\approx \ln(2) \Sigma$ Golomb, numerical optimization leads to $M\approx 0.66794\, \Sigma$ formula. In practice this $M$ has to be rounded to natural number, preferably with a power-of-2 $M=2^m$ what allows to inexpensively directly work on bits.

Denoting $X=\lfloor x/M \rfloor$, its probability distribution is $\textrm{Pr}(X) = a^{MX}\,(1-a^M)$. Average number of bits/value is:

$$h=\sum_{X=0}^\infty (\lg(M)+X+1)\, a^{MX} (1-a^M) = \frac{1}{1-a^M}+\lg(M)$$
Subtracting $h_0=-\lg(1-a)-\frac{a \lg(a)}{1-a}$ entropy of $\textrm{Pr}(x) = a^x\,(1-a)$ geometric distribution, we get $\approx 0.027$ bits/value penalty for larger $\Sigma$. This penalty mainly comes from using uniform distribution instead of geometric for $\textrm{mod}(x,M)$, what could be improved using e.g. rANS here. Restring to power-of-2 $M=2^m$, there appear oscillations around $\approx 0.06$ bits/value penalty - the plots are presented in bottom-right of Fig. \ref{sigma}. Mean penalty is smaller: multiplied by Laplace probability of tail, but for large $\Sigma$ it is not very helpful.
\subsubsection{Direct LSB flush} More practical seems direct flush of $m$ youngest bits, and use e.g. rANS for the remaining bits. Analogously denoting $M=2^m$, we get penalty of such approximation (of uniform distribution for $\textrm{mod}(x,M)$):
$$h=\sum_{X=0}^\infty (\lg(M)-MX\lg(a)-\lg(1-a^M))\, a^{MX} (1-a^M) = $$ $$=\lg(M)-\frac{a^M M \lg(a)}{1-a^M}-\lg(1-a^M)\quad \textrm{bits}$$
Bottom-right of Fig. \ref{sigma} shows $h-h_0$ penalty for $M=2^m$ and $m=1,2,3,4$. For discussed $E=1/300$ we basically can get to $\Sigma \approx 8 $ this way by flushing $\approx \lg(\Sigma/8)$ least significant bits and encoding the remaining with entropy coder, what seems the most practical compromise.

\section{Flexible density quantization}
Having a model of 1D density $\rho:D\to \mathbb{R}^+$ (integrating to 1, usually $D=\mathbb{R}$) e.g. as Laplace or EPD, there remains crucial question of choosing quantization.

A standard choice is uniform quantization as computationally inexpensive, but it might leave improvement opportunities. On the opposite side there is Lloyd-Max algorithm~\cite{lloyd,max} performing costly mean distortion optimization for a chosen number of regions ($N$). It neglects entropy growth which turns important issue - included in considerations here.

There is discussed approach combining their advantages: for a fixed parametric distribution, we would like to automatize inexpensive process of optimized quantization into flexible number of regions $N$, with control of rate and distortion.

For this purpose, let us introduce \textbf{quantization density function} $q:D\to \mathbb{R}^+$, also integrating to 1, intuitively defining how dense local quantization should be. We will optimize it accordingly to assumed density $\rho$. Analogously to cumulative distribution function (CDF), let us define:
\be Q(x)=\int_{-\infty}^x q(x) dx \in [0,1] \ee
We can use it to define  centers of quantization regions for any number of regions $N$ by taking inverse CDF on some a regular lattice, for example:
\be \mathcal{Q}=\{Q^{-1}((i-1/2)/N):i=1,\ldots,N\} \ee
For what encoder needs to perform $Q(x)$, e.g. tabled or interpolated in practical realizations, then perform uniform quantization on $[0,1]$. Decoder analogously needs tabled/interpolated $Q^{-1}$ function:
\be \hat{x}=\lceil N Q(x))\rceil \qquad \tilde{x}=Q^{-1}((\hat{x}-1/2)/N) \ee
Another approach is taking boundaries of quantization regions in the middle between succeeding points of $\mathcal{Q}$  - what is used in evaluations.

Optimization of $q$ for a given $\rho$ can be done for $N\to \infty$ continuous limit, for which we can assume that local distance between quantization nodes in position $x$ is approximately $(N q(x))^{-1}$.

\begin{figure}[t!]
    \centering
        \includegraphics{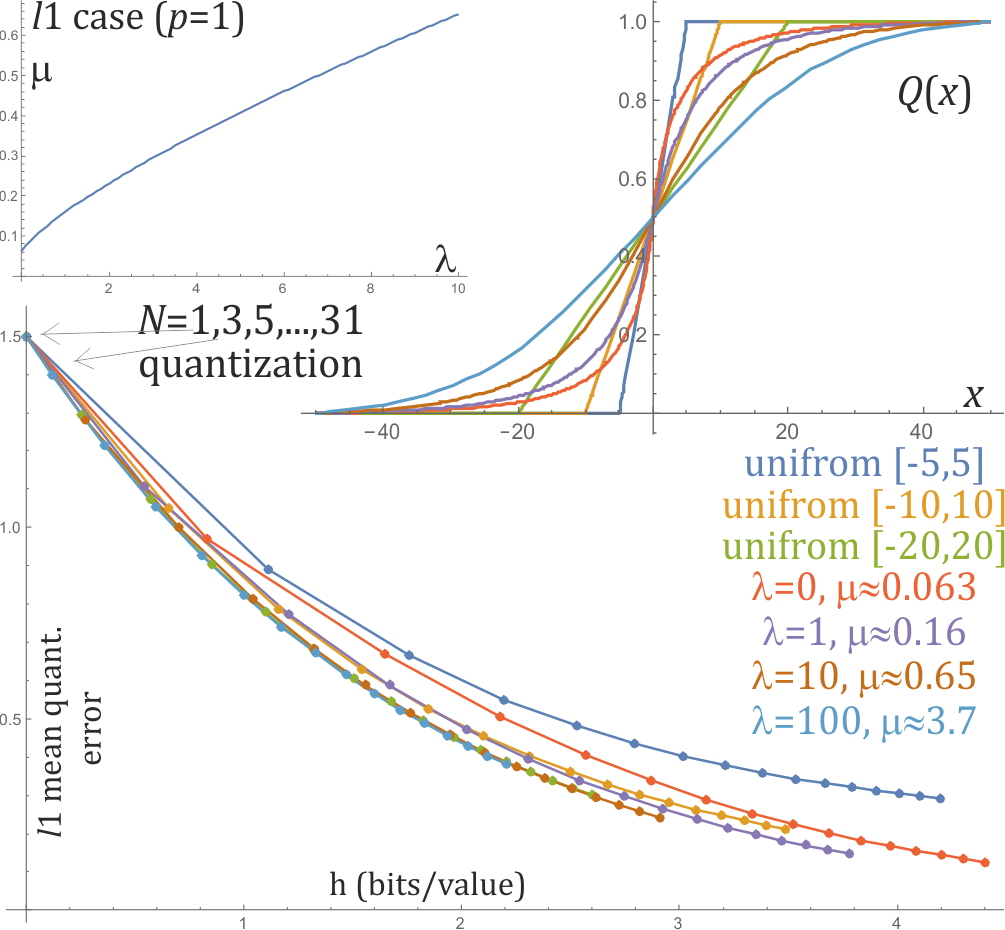}
        \caption{Discussed rate-distortion $l^1$ ($p=1$) optimization of the quantization density: $q=\sqrt{\mu\rho+\lambda^2 \rho^2}-\lambda\rho$, for $(\lambda,\mu)$ satisfying $\int q dx=1$ - such pairs are presented in \textbf{top-left} plot. Their $Q(y)=\int_{-\infty}^{y} q(x) dx$ for some 4 cases are presented in \textbf{top-right} plot, together with such functions providing uniform quantization on ranges (3 different). \textbf{Bottom}: rate-distortion evaluation for all these cases and quantization size $N=1,3,\ldots,31$ (only odd as essentially better than even). We can see that the best ones approach uniform quantization on the largest range $[-20,20]$ (green points).}
       \label{quant}
\end{figure}
\subsection{Distortion  minimizing quantization}
For distortion defined as mean power $p$ of quantization error $|x-\tilde{x}|^p$, e.g. $p=2$ for popular MSE, we can say that mean distortion in position $x$ is proportional to $1/(Nq(x))^p$. Averaging such local distortion over assumed probability distribution $\rho$, we get distortion evaluation:
\be \mathcal{D}(q)\equiv\mathcal{D}=\int_{x\in D} \frac{\rho(x)}{(q(x))^p} dx\qquad \mathcal{D}_N= \frac{\mathcal{D}}{N^p} \ee
where $\mathcal{D}_N$ is approximation for quantization into $N$ regions. To choose the optimal $q$ we can use calculus of variations (e.g. \cite{arnold}): to minimize $\mathcal{D}$ as in the necessary condition for extremum, the first order correction of $\mathcal{D}$ for any (infinitesimal) perturbation $q\to q+\delta q$ has be 0, for $\delta q$ being a function integrating to 0 to maintain $\int q dx=1$ normalization:
$$\forall_{\delta q: \int \delta q\, dx =0}\quad 0=\mathcal{D}(q+\delta q)-\mathcal{D}\approx -p\int \frac{\rho(x)}{(q(x))^{p+1}} \delta q(x) dx$$
It is always zero if $\rho(x)/(q(x))^{p+1}=\textrm{const}$. Otherwise, we could increase $\delta q$ where this fraction is larger, at cost of where it is smaller, getting nonzero variation.

So $\mathcal{D}$ is minimized for $\rho(x)/(q(x))^{p+1}=\textrm{const}$, getting:
\be q(x)=(\rho(x))^{1/(p+1)} /\int (\rho(y))^{1/(p+1)} dx \label{dist}\ee
for normalization to integrate to 1. For MSE we have $p=2$: quantization density $q$ should be increased with cube root of density $\rho$, e.g. twice denser for 8 times larger $\rho$.

While generally we can find $q, Q$ numerically and store in tables for fixed center $\mu=0$ and scale parameter $\sigma=1$ (for shifted and recaled values), for discussed general EPD family (containing e.g. Laplace and Gaussian distribution), we know their analytical formulas as they are just rescaled original distributions:
\be \sigma_q = \sigma/\sqrt[\kappa]{p+1} \qquad q=\rho_{\kappa \mu\sigma_q}\qquad Q=F_{\kappa\mu \sigma_q }\ee

\subsection{Entropy (rate) minimizing quantization}
To calculate asymptotic $N\to \infty$ behavior of entropy (required bits/value rate), probability of quantization region in position $x$ is asymptotically $\rho(x)/(N q(x))$: requiring $\lg((N q(x))/\rho(x))$ bits.
$$ \mathcal{H}_N=\int \rho(x) \lg\left(\frac{N q(x)}{\rho(x)}  \right) dx =\mathcal{H}+\lg(N) $$
\be\textrm{for}\qquad \mathcal{H}=\int \rho(x) \lg\left(\frac{q(x)}{\rho(x)}  \right) dx\ee
what is minus Kullback-Leibler divergence: gets extremum in $q=\rho$ (can be obtained with above calculus of variations), but this time maximal number of bits/value - we would like to get far away from it.

The minimal entropy we could get here is usually zero: by quantization which puts practically entire probability into single region, but it makes no sense from practical perspective - optimizing quantization density to minimize entropy alone rather makes no sense.

\subsection{Rate-distortion optimization}
Distortion optimization alone indeed reduces it for a given quantization size, however, it happens at cost of increased entropy as it leads to more uniform probability distribution among quantization regions than uniform quantization (should concern also e.g. Lloyd-Max).

Hence, in practice we should optimize distortion and entropy together, what can be done using Lagrange multipliers. We have two constraints here, each gets one multiplier: first for normalization $\int q dx=1$ (previously hidden e.g. as $\textrm{const}=\rho/q^{p+1}$). Second for fixed entropy or distortion - while minimizing the other, both these cases are mathematically similar.

Finally, we can just use some two multipliers $\mu, \lambda$, focus on their pairs maintaining normalization $\int q dx=1$, getting (entropy, distortion) pairs hopefully being in minimum (not maximum or saddle). For this purpose we can start with the safe:  distortion optimization case (\ref{dist}) and try to continuously (e.g. numerically) modify it solving ordinary differential equation obtained by treating $\int q dx=1$ as implicit equation.

\subsubsection{$l^1$ quantization error ($p=1$)}
To simplify the solution formula, for $p=1$ case let us choose $\mu, \lambda$ Lagrange multipliers in the following way:
\be\mu\frac{\rho}{q^{p+1}}-2\lambda\frac{\rho}{q}=1\quad\textrm{satisfying}\quad \int q dx=1\ee
For  $\mu, \lambda, \rho, q\geq 0$ we get promising solution:
\be q=\sqrt{\mu\rho+\lambda^2 \rho^2}-\lambda\rho=\frac{\mu\rho }{\sqrt{\mu\rho+\lambda^2 \rho^2}+\lambda\rho}\ee
For $\lambda=0$ we get $q\propto \rho^{1/2}$ as for distortion minimization. For $\mu=0$ we get the problem of entropy optimization case.

Figure \ref{quant} contains such $(\lambda,\mu)$ pairs satisfying $\int q dx=1$, obtained by just optimizing $\mu$ for succeeding $\lambda>0$ on a lattice. It leads to $\mu, \lambda\to \infty$ with fixed asymptotic $\mu/\lambda$. This limit approaches constant $q$ case of uniform quantization, with additionally optimized handling of tails.

\subsubsection{MSE: $l^2$ quantization error ($p=2$)}
For $p=2$ we get degree 3 polynomial instead of 2, which still has analytical solution, but a bit more complex one - we can perform analogous analysis, what is planned for further versions of this articles.

\section{Conclusions and further work}
While often there are uncritically used assumptions of naturally looking distributions, like Laplace or Gaussian, it might be worth testing also e.g. more general families, like EPD discussed here, or heavy tail like stable distributions appearing e.g. in generalized central limit theorem for addition of i.i.d infinite variance variables. For data compression applications, improvements of likelihood can be directly translated to savings in bits/value. Also, while DCT transform decorrelates coefficients, there remain other statistical dependencies like homoscedasticity - their exploitation is computationally more costly, but as discussed can lead to relatively huge savings.

There was also discussed approach for automatic search of flexible quantization, shifting the problem into finding e.g. continuous quantization density function for $N\to \infty$ continuous limit, and then use it for finite $N$. While it can improve distortion alone, together with entropy optimization it seems to lead to nearly uniform quantization - with additional  tail optimization.\\

Discussed flexible quantization approach  needs further work, starting with finishing $p=2$ case and testing for various distributions. It generally brings a question of practicality also of approaches like Lloyd-Max, what requires deeper exploration.

It might be also valuable to try to expand this work into vector quantization. From classical PVQ~\cite{pvq} perspective, there might be considered deformation to reduce distortion - without it is nearly optimal for Laplace distribution, it can be deformed to optimize for Gaussian distribution with uniform on sphere~\cite{pvq1}, we can also use this technique for obtained here deform for $\kappa=1/2$ EPD distribution.

There is also planned further analysis and testing of context dependent methods from \cite{param, param1} for inexpensive prediction of parameters e.g. $\mu,\sigma$ of $\kappa=1/2$ EPD distribution, testing statistical dependencies for coefficients inside $8\times 8$ DCT block and between neighbors, e.g. using observations for larger $16\times 16$ block treated as four $8\times 8$ sub-blocks and predicting for one of them from already decoded three.

Further large topic to consider is optimizing transforms, especially color as discussed in \cite{param1}, maybe also entire DCT-like transform, combination with chroma subsampling for perceptual evaluations.

\bibliographystyle{IEEEtran}
\bibliography{cites}
\end{document}